\documentclass[twocolumn,prb,amsmath,amssymb,floatfix,superscriptaddress]{revtex4-2} 
\usepackage{amsmath}
\usepackage{amssymb}
\usepackage{graphicx}
\usepackage{bm}
\usepackage{color}
\usepackage{hyperref}
\usepackage{graphicx}
\usepackage{upgreek}
\usepackage{scalerel}
\usepackage{multirow}
 
 \usepackage[bb=dsserif]{mathalpha}
 
\usepackage{pgffor}
\usepackage{pdfpages}
\usepackage{mathdots}
\usepackage{float}
\usepackage{silence}
\usepackage{physics} 
\usepackage{lscape}   
\usepackage{color, colortbl}
\usepackage[table]{xcolor}
\usepackage{comment}  
\makeatletter
\AtBeginDocument{\let\LS@rot\@undefined}
\makeatother

\begin{document}

\title{Non-Hermitian zero-energy pinning of Andreev and Majorana bound states
 in superconductor-semiconductor systems}

\author{Jorge Cayao}
\affiliation{Department of Physics and Astronomy, Uppsala University, Box 516, S-751 20 Uppsala, Sweden}
\date{\today}

\begin{abstract}
The emergence of Majorana bound states in finite length superconductor-semiconductor hybrid systems has been predicted to occur in the form of oscillatory energy levels  with parity crossings around zero energy. Each zero-energy crossing is expected to produce a quantized zero-bias conductance peak but several studies have reported   conductance peaks pinned at zero energy over a range of Zeeman fields, whose origin, however, is not clear.  In this work we consider superconducting systems with spin-orbit coupling under a Zeeman field and demonstrate that non-Hermitian effects, due to coupling to ferromagnet leads, induce  zero-energy pinning of Majorana and trivial Andreev bound states. We find that this zero-energy pinning effect occurs due to the formation of non-Hermitian spectral degeneracies known as exceptional points, whose emergence can be controlled by the interplay of non-Hermiticity, the applied Zeeman field, and chemical potentials.  Moreover, depending on the non-Hermitian spatial profile, we find that  non-Hermiticity changes the single point Hermitian topological phase transition into a flattened zero energy line bounded by exceptional points from multiple low energy levels. This seemingly innocent change notably enables a gap closing well below the Hermitian topological phase transition, which can be in principle simpler to achieve. Furthermore, we reveal that the energy gaps separating Majorana and trivial Andreev  bound states from the quasicontinuum  remain robust for the values that give rise to the zero-energy pinning effect. While reasonable values of non-Hermiticity can be indeed beneficial, very strong non-Hermitian effects can be detrimental as it might destroy superconductivity. Our findings can  be therefore   useful for understanding the zero-energy pinning of trivial and topological states in Majorana devices.
\end{abstract}
\maketitle

\section{Introduction}
The search of Majorana bound states (MBSs) in topological superconductors has become one of the central topics in condensed matter due to their potential for robust quantum computing \cite{sarma2015majorana,beenakker2019search,aguado2020majorana,aguado2020perspective}. While topological superconductivity and MBSs were initially predicted in intrinsic spin-triplet $p$-wave superconductors, their physical realization has been mostly pursued in superconductor-semiconductor hybrids \cite{tanaka2011symmetry,sato2016majorana,sato2017topological,Aguadoreview17,lutchyn2018majorana,zhang2019next,frolov2019quest,prada2019andreev,flensberg2021engineered,Marra_2022,tanaka2024theory}.   In this hybrid setup, an applied magnetic field  induces a topological phase transition, after which MBSs emerge as edge states with their energies oscillating around zero energy in the form of parity crossings \cite{PhysRevB.86.180503,PhysRevB.86.220506,PhysRevB.87.024515,PhysRevB.96.205425}. In sufficiently long systems, MBSs acquire zero energy, a property that has been  explored in conductance experiments but their Majorana interpretation is still puzzling \cite{prada2019andreev}. 

On of the main issues in conductance experiments is that the reported zero bias conductance peaks (ZBCPs) are often pinned at zero energy over a range of magnetic fields \cite{Albrecht16,PhysRevLett.123.107703,zhang2021large,valentini2020nontopological}, which, however, do not have a unique explanation, see also Refs.\,\cite{prada2019andreev,zhang2019next}. In fact, ZBCPs can form due to MBSs 
\cite{TK95,KT96,Kashiwaya_RPP,PhysRevLett.98.237002,PhysRevLett.103.237001,PhysRevB.82.180516}  but also due to topologically trivial Andreev bound states (TABSs) \cite{Bagrets:PRL12,Pikulin2012A,PhysRevB.86.100503,PhysRevB.91.024514,PhysRevB.86.180503,PhysRevB.98.245407,PhysRevLett.123.117001,DasSarma2021Disorder,PhysRevB.104.L020501,PhysRevB.104.134507,PhysRevB.107.184509,baldo2023zero,PhysRevB.107.184519}, with both types of states susceptible to a zero-energy pinning effect. In the case of MBSs, it has been shown that electronic interactions \cite{dominguez2017zero,escribano2018interaction} and dissipation \cite{JorgeEPs,avila2019non,cayao2023nonhermitian,ghosh2024majorana} are possible mechanisms for inducing a zero-energy pinning, while  
 very strong SOC \cite{PhysRevB.91.024514,PhysRevB.93.245404,PhysRevB.102.245431} or multiple bands \cite{PhysRevB.100.125407} are needed for the zero-energy pinning of TABSs. While disorder and multiple bands are intrinsic in superconductor-semiconductor  system and thus difficult to control, the fabrication of cleaner and lower dimensional samples can in principle mitigate these effects. However, the effect of dissipation, naturally appearing   when attaching normal reservoirs in transport measurements \cite{zhang2019next,prada2019andreev}, cannot be avoided, thus highlighting its relevance on the formation of MBSs and TABSs.

The effect of dissipation has also been shown to have profound consequences beyond its role as a zero-energy pinning mechanism of MBSs. Indeed, dissipation renders the system non-Hermitian and enables entirely novel topological phases that do not exist in the Hermitian realm \cite{doi:10.1080/00018732.2021.1876991,OS23} as well as intriguing bulk Fermi arcs \cite{kozii2017,PhysRevB.97.041203,Zhou_2018,PhysRevB.98.035141,doi:10.7566/JPSCP.30.011098,PhysRevB.107.104515} and unusual transport properties \cite{JorgeEPs,PhysRevB.98.155430,PhysRevB.109.165402,cayao2023nonhermitian,li2023anomalous,PhysRevB.100.245416,PhysRevB.107.035408,PhysRevB.105.155418,cayaominimalkitaev}. These exotic phenomena originate from the emergence of singular points in parameter space known as \textit{exceptional points (EPs)}, defined as points where two or more eigenvalues (and their respective eigenfunctions) coalesce \cite{TKato, heiss2004exceptional, berry2004physics, Heiss_2012, PhysRevLett.86.787, PhysRevLett.103.134101, PhysRevLett.104.153601, gao2015observation, doppler2016dynamically,PhysRevB.99.121101}. Despite the unavoidable presence of dissipation in  superconductor-semiconductor systems, the impact of EPs on superconducting systems hosting simultaneously TABSs and MBSs still remains an open question.  Specially, superconducting systems with an homogeneous dissipation in space remain unexplored. By characterizing the role of dissipation, it would be possible to advance the understanding of Majorana devices.

\begin{figure}[!t]
	\centering
	\includegraphics[width=0.495\textwidth]{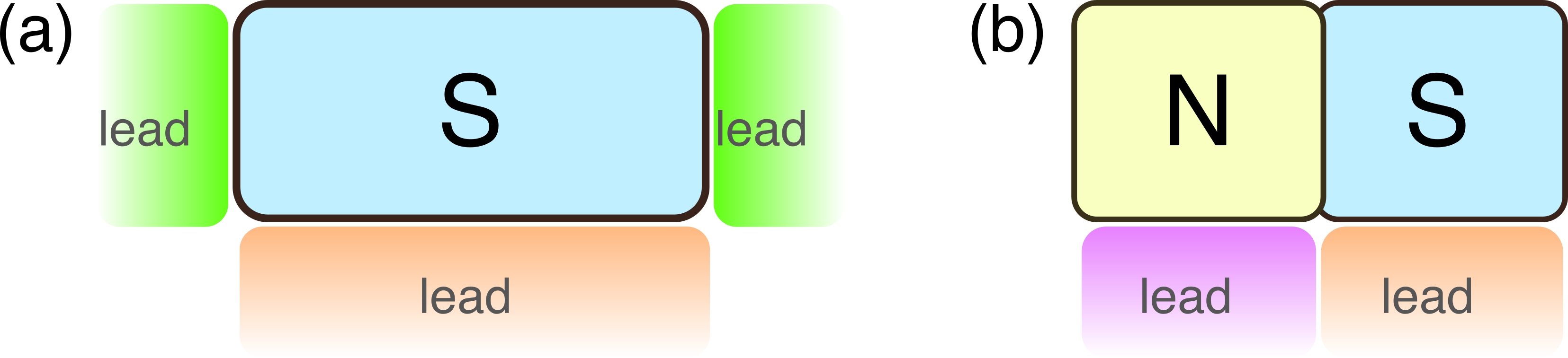} 
	\caption{Sketch of the studied non-Hermitian superconducting systems, where non-Hermiticity appears due to coupling to normal reservoirs or leads. (a) A Rashba superconductor (S in cyan) is coupled to three  leads  in the normal state (green and orange boxes), which can be  ferromagnetic. While the green leads are only coupled to the left and right sides of S, the orange lead is coupled to the \emph{entire} S in an homogeneous fashion. (b) A NS junction  with the \emph{entire} N and S regions coupled to ferromagnet leads (magenta and orange boxes).}
	\label{fig1}
\end{figure}

In this work, we consider two one-dimensional (1D) superconducting systems with Rashba spin-orbit coupling (SOC) under a Zeeman field and study the response of their low-energy spectrum to non-Hermitian effects  when coupling  to ferromagnet reservoirs or leads. In particular,  we explore finite length non-Hermitian systems including a superconductor and  a normal-superconductor (NS) junction coupled to ferromagnet leads as in Fig.\,\ref{fig1}, permitting  us to inspect MBSs and TABSs at the same footing. In general we demonstrate  that  non-Hermiticity induces a zero-energy pinning of MBSs and TABSs, an effect that emerges as lines of zero real energy whose ends mark the formation of exceptional points. This zero-energy pinning effect can be controlled by the interplay of non-Hermiticity and the system parameters, such as the Zeeman field and chemical potentials. By increasing non-Hermiticity, however, the evolution of the zero-energy pinning effect of TABSs exhibits a different behaviour than that of MBSs. We also discover that an homogeneous non-Hermiticity in the superconductor transforms the Hermitian topological phase transition occurring at a single point into a zero-energy line with exceptional points, which then gives rise to a gap closing at much lower Zeeman fields. Furthermore, we show that the values of non-Hermiticity causing the zero-energy pinning effect do not affect the energy gap separating TABSs and MBSs from the quasicontinuum, revealing the beneficial effect of low dissipation.  Very strong non-Hermiticity, however, can induce a zero-energy pinning of the energy gaps and also of high energy levels, which can be detrimental for Majorana applications. Our results therefore demonstrate that dissipation-induced non-Hermiticity is a potential mechanism to produce zero-energy pinning of trivial and topological states in superconductor-semiconductor systems.

The remainder of this work is organized as follows. In  Section \ref{section2} we 
discuss the effective non-Hermitian Hamiltonian for a superconductor with SOC coupled to ferromagnet leads. In Section \ref{section3} we show the impact of non-Hermiticity on the low-energy spectrum of a finite length superconductor with homogeneous pair potential, while in Section \ref{section4} we address NS junctions.  Finally, in Section \ref{section5}, we present our conclusions.

\section{Effective non-Hermitian model}
\label{section2}
We are interested in exploring the impact of non-Hermiticity on trivial and topological zero-energy states in superconductor-semiconductor hybrids. For this purpose we consider a 1D  superconductor with Rashba SOC, which captures the main properties of superconductor-semiconductor hybrids \cite{tanaka2024theory}, and coupled it to normal reservoirs such that the total system is open and described by an effective non-Hermitian Hamiltonian, see Fig.\,\ref{fig1}. In particular,  the 1D open system can be  modelled by an effective  Hamiltonian given by
\begin{equation}
\label{NHH}
H_{\rm eff}=H_{\rm S}+\Sigma^{r}(\omega=0)\,,
\end{equation}
where $H_{\rm S}$ is the Hermitian Hamiltonian  describing the closed superconductor with SOC, while $\Sigma^{r}(\omega=0)$ is the zero-frequency retarded self-energy that  incorporates the non-Hermitian effects due to  coupling to normal reservoirs. Even though the self-energy is in general frequency dependent, it can be  approximated by its zero-frequency version $\Sigma^{r}(\omega=0)$ in the wide-band limit \cite{datta1997electronic}, and its form will be explicitly given below. This wide-band limit has also been shown to induce interesting  non-Hermitian effects in bulk setups \cite{PhysRevB.105.094502,PhysRevB.107.104515,cayao2023exceptional}.

We  consider that the Hermitian superconductor with SOC is under the presence of a magnetic field since we are interested in obtaining MBSs \cite{tanaka2024theory}. Thus, the Hermitian superconductor  is modelled by a spinful one-dimensional (1D) tight-binding chain given by
\begin{equation}
\label{modelTBNW}
\begin{split}
H_\textrm{S}&=\varepsilon\sum_{\sigma n}c_{\sigma n}^\dagger c^{\phantom{\dagger}}_{\sigma n} 
-\mathop{\sum_{\langle n,n'\rangle}}_{\sigma}t\, c_{\sigma n'}^\dagger c^{\phantom{\dagger}}_{\sigma n} 
\\
&-i\mathop{\sum_{\langle n,n'\rangle}}_{\sigma,\sigma'}t^{\mathrm{SOC}\phantom{\dagger}}_{n'-n}c_{\sigma' n'}^\dagger \sigma^{y}_{\sigma'\sigma}c^{\phantom{\dagger}}_{\sigma n}\\
&+\sum_{\sigma,\sigma' n}B\,c_{\sigma' n}^\dagger \sigma^{x}_{\sigma'\sigma}c^{\phantom{\dagger}}_{\sigma n}+
\sum_{\sigma n}\Delta_{n}\, c_{\sigma n}^\dagger c^\dagger_{\bar\sigma n}+\mathrm{H.c}\,,
\end{split}
\end{equation} 
where $c_{\sigma n}$ destroys a fermionic state with spin $\sigma$ at site $n$ that runs over all the M lattice sites of the system of length   $L=Ma$, with  $a$ being the lattice spacing. Moreover, here $\varepsilon=2t-\mu$ is the onsite energy, $\mu$ is the chemical potential that determines the filling,  $t=\hbar^2/(2m a^2)$ is the hopping,   $m$ is the effective mass, and $\langle n,n'\rangle$ indicates hopping between nearest neighbor sites. Moreover, $t^\mathrm{SOC}_{\pm 1}=\pm \alpha_\mathrm{R}/(2a)$ is the the SOC hopping, where $\alpha_\mathrm{R}$ is the SOC strength that defines a SOC length given by $\ell_\mathrm{SOC}=\hbar^2/(m \alpha_\mathrm{R})$,  $B=g\mu_B \mathcal{B}/2$ is the Zeeman field due to an external magnetic field $\mathcal{B}$ along the wire and perpendicular to the SOC axis, while $g$ is the g-factor.  Furthermore,  $\Delta_{n}$ represents the space dependent proximity-induced spin-singlet $s$-wave pair potential from the superconductor into the semiconductor.   When the pair potential is homogenous, namely, $\Delta_{n}=\Delta$, then  the Hermitian system is a uniform superconductor. On the contrary, when finite regions have $\Delta_{n}=0$ and $\Delta_{n}=\Delta$, we refer to such regions as to normal (N) and superconducting (S) regions, respectively. In this case, we   have a NS junction, which will be also studied here because these junctions naturally host trivial zero-energy states \cite{PhysRevB.91.024514,PhysRevB.104.L020501}.

As already mentioned above, the effect of the normal reservoirs, here referred to  as normal leads, is taken into account in the form of a zero-frequency retarded self-energy, which is commonly done for studying transport \cite{PhysRevB.54.7366,datta1997electronic}. Motivated by the fact that distinct normal leads are usually coupled   to superconductor-semiconductor systems for carrying out transport experiments \cite{lutchyn2018majorana,prada2019andreev,zhang2019next,flensberg2021engineered}, here we consider three distinct ferromagnet  leads  as depicted in Fig.\,\ref{fig1}.  Furthermore, we note that, even though the self-energy has in general real (Re) and imaginary (Im) parts, only its Im part induces non-Hermitian effects which will be studied here; its Re part renormalizes the diagonal entries of $H_{\rm S}$. Thus, the total self-energy can be written as  
\begin{equation}
\label{selfEn}
\Sigma^{r}(\omega=0)=\Sigma^{r}_{\rm L}+\Sigma^{r}_{\rm R}+\Sigma^{r}_{\rm X}\,,
\end{equation}
where
\begin{equation}
\label{SelfEnergies}
\begin{split}
\Sigma^{r}_{\rm L}&=-i\sum_{\sigma}\Gamma_{1\sigma}c_{\sigma 1}^\dagger c_{\sigma 1}\,,\\
\Sigma^{r}_{\rm R}&=-i\sum_{\sigma}\Gamma_{M\sigma}c_{\sigma M}^\dagger c_{\sigma M}\,,\\
\Sigma^{r}_{\rm X}&=-i\sum_{\sigma n}\Gamma_{n\sigma}c_{\sigma n}^\dagger c_{\sigma n}\,,\\
\end{split}
\end{equation}
model the coupling of the Hermitian system to the left, right, and middle ferromagnet leads, respectively. See Ref.\,\cite{PhysRevB.105.094502} for details on the derivation of Eqs.\,(\ref{SelfEnergies}). Here, $\Sigma^{r}_{\rm L(R)}$ has finite values  at the first (last) site, where $\Gamma_{\alpha\sigma}$ characterizes the coupling of the first ($\alpha=1$) and last site ($\alpha=M$) to the leads. Similarly, $\Sigma^{r}_{\rm X}$ is the self-energy due to coupling  a lead (or leads) to the entire system   with $\Gamma_{\alpha\sigma}$ and $\alpha\in(1,M)$, see orange lead in Fig.\,\ref{fig1}(a)  and also magenta/orange lead in Fig.\,\ref{fig1}(b).  The couplings can be written as \cite{PhysRevB.105.094502} $\Gamma_{\alpha\sigma}=\pi|\tau_{\alpha}|^{2}\rho_{\rm \alpha}^{\sigma}$, being $\tau_{\alpha}$ the hopping   into the lead $\alpha=L,R,X$ from the Hermitian superconductor and $\rho_{\rm \alpha}^{\sigma}$  the surface  density of states of the lead $\alpha$ for spin $\sigma=\uparrow, \downarrow$. 

Before going further we note that both $H_{\rm S}$ and $\Sigma^{r}$ in  Eq.\,(\ref{modelTBNW}) and Eq.\,(\ref{selfEn}), respectively, are given in terms of creation and annihilation operators. In this regard,  we can write  Eq.\,(\ref{NHH}) in 
Nambu space $(c_{n\sigma},c_{n\sigma}^{\dagger})$, which implies that the non-Hermitian effective Hamiltonian can be treated as a matrix in real space whose dimensions are defined by the number of lattice sites. Moreover, it is worth noting that the effective Hamiltonian has particle-hole symmetry given by $H_{\rm eff}=-\hat{C}^{-1}H^{*}_{\rm eff}\hat{C}$, where $\hat{C}=\sigma_{0}\tau_{x}C$ and $C$ is the complex conjugation operation \cite{pikulin2012topological,PhysRevB.87.235421,ioselevich2013tunneling,RevModPhys.87.1037,JorgeEPs}; this symmetry dictates that the eigenvalues of $H_{\rm eff}$ come in pairs as $E_{n}$ and $-E_{n}^{*}$. Furthermore, given that the spectrum of the effective Hamiltonian in Eq.\,(\ref{modelTBNW}) corresponds to the poles of a retarded Green's function, the poles (and hence the eigenvalues) reside in the the lower complex energy half-plane. This, combined with the particle-hole symmetry imposes a real spectrum that is symmetric around zero, while an imaginary part that is negative but not symmetric around zero, as we will see below.

We are interested in  exploring the impact of non-Hermiticity due to normal leads modelled by Eqs.\,(\ref{SelfEnergies}) on the emergence of MBSs and TABSs in $H_{\rm S}$  modelled by Eq.\,(\ref{modelTBNW}). While the formation of MBSs \cite{tanaka2024theory} and TABSs \cite{PhysRevB.91.024514,PhysRevLett.123.117001,PhysRevB.107.184519}  appear in closes systems and do not require the presence of leads, having closed systems coupled to leads as in Fig.\,\ref{fig1} provides an interesting scenario to explore the impact of non-Hermiticity due to the leads on MBSs and TABSs. To address these question, we   consider realistic  parameters, with $\alpha_{\rm R}=20$\,meVnm and $\Delta=0.25$\,meV, according to experimental values reported for InSb and InAs semiconductor nanowires and Nb and Al superconductors \cite{lutchyn2018majorana}. Moreover, we take the lattice spacing of  $a=10$nm and analyze systems of realistic lengths. Taking these realistic parameters, we investigate the formation of MBSss and TZABSs under non-Hermiticity by using Eq.\,(\ref{NHH}).

\section{Non-Hermitian Rashba superconductor}
\label{section3}
We start by analyzing  the impact of non-Hermiticity on the formation of MBSs in  the non-Hermitian superconductor  modelled by Eq.\,(\ref{modelTBNW}), with an homogeneous pair potential $\Delta_{n}=\Delta$ and a constant self-energy  all over the system that is only given by $\Sigma_{\rm X}$ from Eqs.\,(\ref{SelfEnergies}). For obvious reasons here we denote X$=$S in order to highlight that the lead is coupled to the entire superconductor S, as indicated by orange lead in Fig.\,\ref{fig1}(a). Since the self-energy is taken to be constant in space all over the superconductor but spin dependent, we consider that the coupling strengths are given by $\Gamma_{n\sigma}=\Gamma_{{\rm S}\sigma}$, see Eqs.\,(\ref{SelfEnergies}). To investigate the emergence of  MBSs under non-Hermiticity, we calculate the energy spectrum of the respective effective non-Hermitian given by Eq.\,(\ref{modelTBNW}). Since the system is non-Hermitian, its spectrum becomes complex:  the real part represents the physical energy of quasiparticles, while the inverse of the imaginary part determines their lifetimes $\hbar/{\rm Im}(E_{n})$, see Ref.\,\cite{datta1997electronic}. When $\Gamma_{{\rm S}\uparrow}=\Gamma_{{\rm S}\downarrow}$, we find that all the eigenvalues acquire the same imaginary part, whose inverse gives the same lifetime for all the eigenvalues.  The situation is, however, distinct when there is an asymmetry in the couplings, namely, $\Gamma_{{\rm S}\uparrow}\neq\Gamma_{{\rm S}\downarrow}$, which we discuss next.

\begin{figure}[!t]
	\centering
	\includegraphics[width=0.495\textwidth]{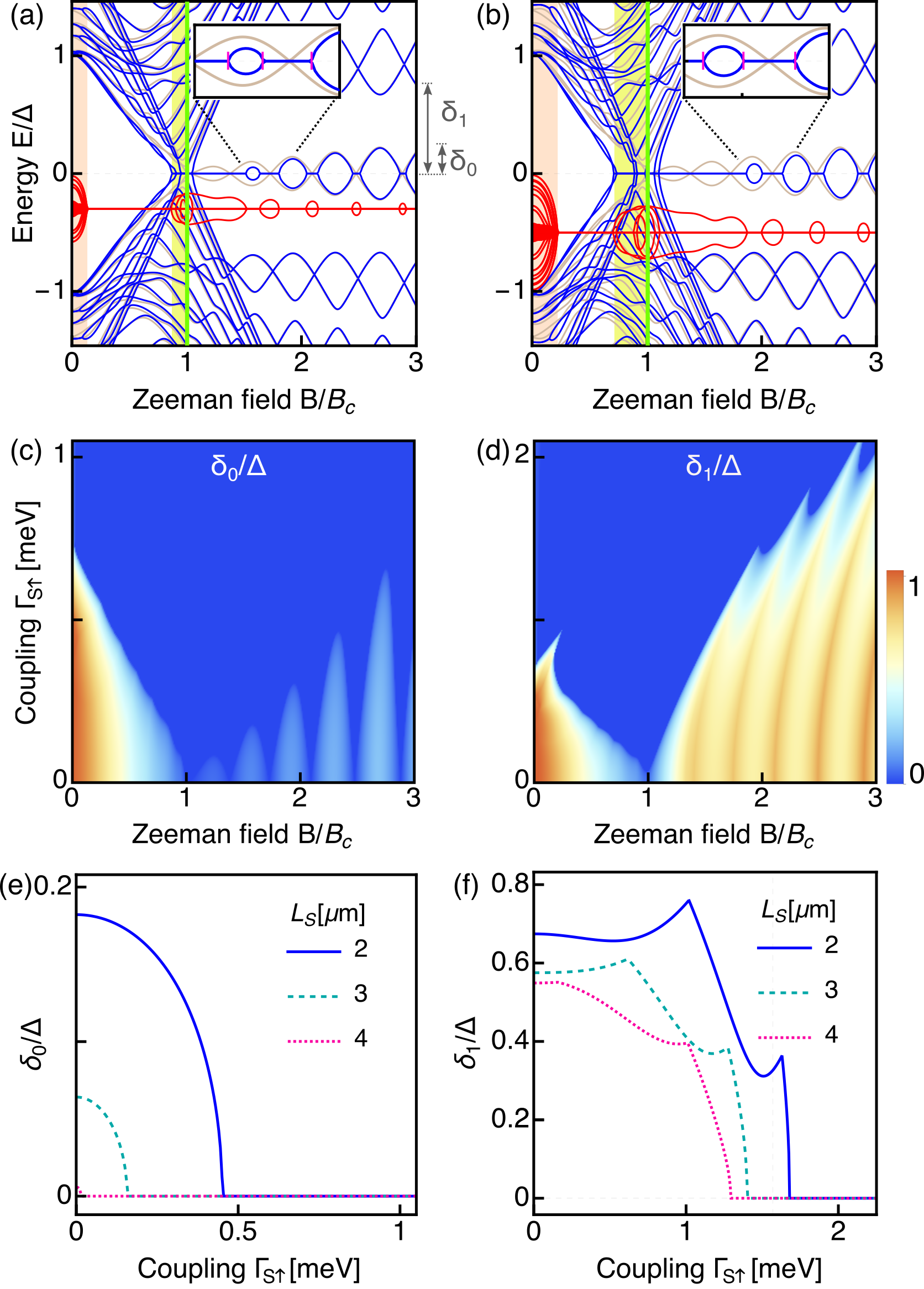} 
	\caption{(a,b) Real (blue) and imaginary (red) energy spectrum   of a finite non-Hermitian Rashba superconductor as a function of the Zeeman field $B$ at two distinct values of homogeneous non-Hermiticity  $\Gamma_{{\rm S}\uparrow}=0.15$meV and $0.25$meV. Green vertical line marks the Hermitian topological phase transition at $B=B_{\rm c}$, while the yellow shaded region indicates its modification due to non-Hermiticity. Brown curves in (a,b) correspond to the eigenvalues without non-Hermiticity. The insets in (a,b) show zoom-in regions of zero-energy lines between EPs (magenta marks).  (c,d) Lowest  and first excited real positive energies  $\delta_{0,1}$  as a function of the Zeeman field $B$ and coupling $\Gamma_{{\rm S}\uparrow}$, with the blue region indicating their vanishing values; note the larger $y$-axis in (d) for $\delta_{1}$. (e,f) $\delta_{0,1}$ as a function of $\Gamma_{{\rm S}\uparrow}$ for distinct values of the length of the superconductor $L_{\rm S}$ at $B=2.3B_{\rm c}$ of (a,b). Parameters: $\alpha_{\rm R}=20$meVnm, $\mu_{\rm S}=0.5$\,meV, $L_{\rm S}=2\mu$m, $\Gamma_{\rm S\downarrow}=0$.}
	\label{fig2}
\end{figure}

In Fig.\,\ref{fig2}(a,b) we present the complex energy spectrum as a function of the Zeeman field $B$ for  $\Gamma_{{\rm S}\uparrow}=0.15$meV and $\Gamma_{{\rm S}\uparrow}=0.25$meV, both at $\Gamma_{{\rm S}\downarrow}=0$ for a finite non-Hermitian Rashba superconductor with $L_{\rm S}=2\mu$m. Here, the real (Re) and imaginary (Im) eigenvalues are depicted in blue and red, respectively. For completeness,  the eigenvalues in the Hermitian regime, having only Re parts and developing loops around zero energy  with zero-energy parity crossings, are shown in brown.   The overall Zeeman dependence of the Re part of the non-Hermitian spectrum roughly follows its Hermitian counterpart, with a symmetric profile around zero energy due to particle-hole symmetry (Sec.\,\ref{section2}), but exhibits some important changes. At very low Zeeman fields, the Re part of the spectrum is gapped and the positive (negative) eigenvalues remain degenerate for a short range of  Zeeman fields, which is, however, different for distinct energy levels, see shaded red region in Fig.\,\ref{fig2}(a,b). The ends of such degenerate Re eigenvalues mark points that are accompanied by the merging of the spin split Im parts, which  signals the emergence of  non-Hermitian degeneracies known as exceptional points (EPs); at these EPs we have verified that the associated eigenvectors coalesce, as expected for EPs \cite{doi:10.1080/00018732.2021.1876991,RevModPhys.93.015005}. Note that having split Im parts means that the associated lifetimes are distinct; their negative  and non symmetric values around zero stem from particle-hole symmetry and causality discussed in  Sec.\,\ref{section2}.  These EPs at finite Re energies have been shown to appear in   bulk Rashba semiconductors as a unique   effect due to the interplay of non-Hermiticity and SOC \cite{cayao2023exceptional} but do not depend on  superconductivity.   

As the Zeeman field increases, the lowest part of the Re spectrum reduces and, notably,  develops a flattened gap closing feature for a range of Zeeman fields around $B=B_{\rm c}$,  where $B_{\rm c}=\sqrt{\mu^{2}+\Delta^{2}}$ marks the Hermitian topological phase transition (vertical green line) after which MBSs emerge \cite{tanaka2024theory}, see yellow shaded region in Fig.\,\ref{fig2}(a,b).  The closing of the Re energy gap acquiring zero energy can be estimated from the bulk Hamiltonian  \cite{PhysRevB.107.104515}, which, at $\mu=0$, is   bounded by $B^{\pm}_{*}=\Delta\pm\gamma$, where $\gamma=(\Gamma_{{\rm S}\uparrow}-\Gamma_{{\rm S}\downarrow})/2$: it evident that non-Hermiticity causes a substantial lower Zeeman field $B^{-}_{*}$ compared to the Hermitian topological phase transition at $B_{\rm c} =\Delta$ when $\mu=0$. The non-Hermitian gap closing is initially  formed by the two lowest  energy levels, which, after an EP transition, stick at zero Re energy for a range of $B$  that is distinct for each  level; the respective Im parts develop loops within EPs, revealing the acquisition of distinct lifetimes. The EPs occur here at zero energy between positive and negative energy levels, which is  distinct to the EPs  discussed in previous paragraph happening between positive energy levels with distinct spin. 

The number of energy levels undergoing  EP transitions at distinct $B$ around $B_{c}$, which also feature a gap closing, can increase depending on how strong is  non-Hermiticity, see  Fig.\,\ref{fig2}(a,b). However, only the lowest (positive and its negative counterpart) energy level remains at zero Re energy as $B$   increases above $B$ after the first gap closing. Interestingly, the Hermitian parity crossings, corresponding to the oscillating energies of MBSs, become pinned at zero Re energy, see Fig.\,\ref{fig2}(a,b).  The ends of the zero Re energies around the parity crossings mark the emergence of EPs, which then determine the effect we refer to as zero-energy pinning; the zero energy Re lines between EPs is more visible in the insets of Fig.\,\ref{fig2}(a,b), where the magneta short lines mark the EPs.  Inside the zero Re energy lines, the Im parts form loops which coalesce at the EPs: this shows that, in an open system,  MBSs acquire a physical   energy equal to zero and distinct imaginary parts that signal their different lifetimes.  While at weak non-Hermiticity only few parity crossings exhibit the zero-energy pinning effect, large non-Hermiticity can also induce  a zero-energy pinning to the zero-energy loops. It is worth noting that the parity crossings, as well as the zero-energy loops, at smaller $B$ are more susceptible to the impact of non-Hermiticity. Thus,   non-Hermiticity is able to reduce the value of Zeeman fields at which the gap closing occurs and also promotes a zero-energy pinning of MBSs.

To gain further insights on the role of non-Hermiticity on MBSs and the gap that protects them from the quasicontinuum, in Fig.\,\ref{fig2}(c,d) we plot the lowest and first excited Re energy levels denoted by $\delta_{0,1}$ in Fig.\,\ref{fig2}(a) as a function of the Zeeman field $B$ and coupling $\Gamma_{{\rm S}\uparrow}$. For obvious reasons, the quantities $\delta_{0,1}$ can be interpreted to be the Majorana energy and the topological gap, respectively. The blue color in  Fig.\,\ref{fig2}(c,d) indicates $\delta_{0,1}=0$, which is achieved much faster for the Majorana energy $\delta_{0}$ than for the topological gap $\delta_{1}$, see that the $y$-axis in (b) runs over a larger values of $\Gamma_{{\rm S}\uparrow}$. It is fair to say, however, that, although non-Hermiticity is indeed beneficial for inducing a zero-energy pinning of MBSs, very strong non-Hermiticity here might be detrimental as it destroys the topological gap [Fig.\,\ref{fig2}(c,d)]. The beneficial and detrimental effects of non-Hermiticity remain even when having longer systems, as shown in  Fig.\,\ref{fig2}(e,f) where we plot $\delta_{0,1}$ as a function of $\Gamma_{{\rm S}\uparrow}$ at $B=2.3B_{\rm c}$ for distinct $L_{\rm S}$. Short and long systems require weak non-Hermiticity to   achieve zero-energy MBSs  which, interestingly,  are much lower than that   needed to  destroy the topological gap.  Therefore, non-Hermiticity due to an homogeneous coupling to ferromagnet leads can be useful to engineer zero-energy MBSs with a well-defined topological gap.

\begin{figure}[!t]
	\centering
	\includegraphics[width=0.495\textwidth]{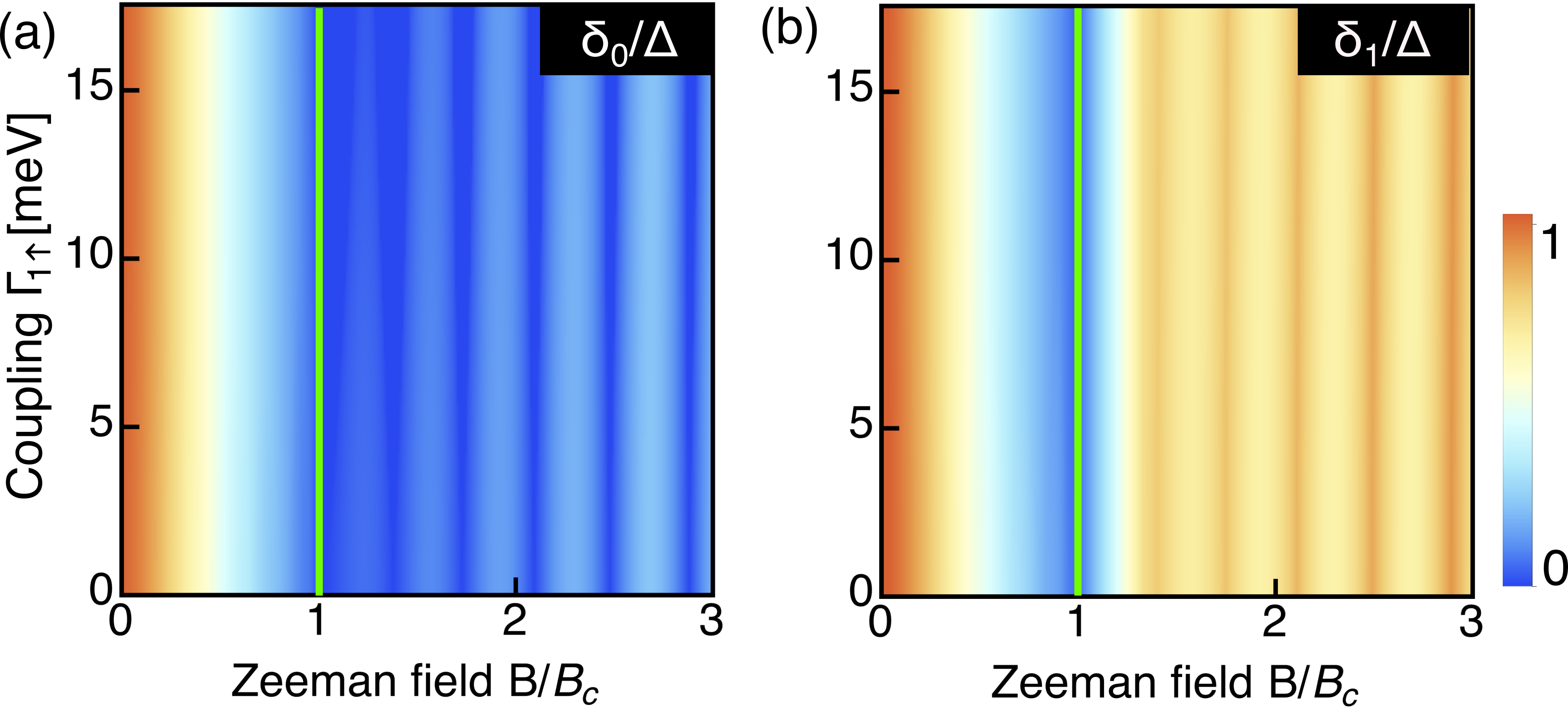} 
	\caption{(a,b) Lowest and first excited  positive real energies $\delta_{0,1}$ of a non-Hermitian Rashba superconductor	as a function of  the Zeeman field $B$ and non-Hermiticity only in the first site with $\Gamma_{1\uparrow}=\Gamma_{1\downarrow}$.   The green vertical line marks the Hermitian topological phase transition $B=B_{\rm c}$. Parameters: $\alpha_{\rm R}=20$meVnm, $\mu_{\rm S}=0.5$\,meV, $L_{\rm S}=2\mu$m.}
	\label{fig3}
\end{figure}

Before going further, we also discuss the impact of non-Hermiticity on the 
Re lowest and first excited energies $\delta_{0,1}$  when the Rashba superconductor is  only coupled to a normal lead on the left (or right) side. In this situation, the effective non-Hermitian Hamiltonian contains a self-energy given by $\Sigma_{\rm L(R)}$ in Eqs.\,(\ref{SelfEnergies}), which only adds the negative imaginary contribution to the first (last) site determined by $\Gamma_{\rm 1(N)\sigma}$ , unlike the case discussed above for an homogeneous non-Hermitian profile of the couplings. For simplicity, we consider that only the left side is coupled to a normal lead such that $\Gamma_{1\uparrow}=\Gamma_{1\downarrow}$ and in Fig.\,\ref{fig3}(a,b)  present $\delta_{0,1}$ as a function of the Zeeman field $B$ and coupling $\Gamma_{1\uparrow}$. Here, the blue color indicates $\delta_{0,1}=0$.  We first observe that  the lowest positive energy level $\delta_{0}$ does not reach zero value before $B_{\rm c}$, even when non-Hermiticity greatly surpasses   the common energy scales of the system such as pair potential and chemical potential. For $B>B_{\rm c}$, we find that $\delta_{0}$, characterizing the Majorana energy, becomes zero  at finite $\Gamma_{1\uparrow}$, as a result of the formation of EPs which then connect zero-energy lines around the zero-energy parity crossings, in the same fashion as found in Fig.\,\ref{fig3}(a-d). This zero-energy pinning of MBSs, however, is visible when non-Hermiticity  $\Gamma_{1\sigma}$ takes very large values, see Fig.\,\ref{fig3}(a).  At these large values on non-Hermiticity the first excited positive energy $\delta_{1}$ only becomes zero at $B=B_{c}$, signaling a single point gap closing  that is different to the case with homogeneous non-Hermiticity of Fig.\,\ref{fig2} but   similar to the Hermitian topological phase transition. Surprisingly,  $\delta_{1}$ maintains a robust    finite value in the topological phase $B>B_{c}$, se seen in Fig.\,\ref{fig3}(b).

We have therefore shown that the effect of non-Hermiticity can be beneficial for stabilizing MBSs without destroying the topological gap. Moreover, depending on the non-Hermitian profile, it is possible to induce a topological phase transition at much lower Zeeman fields, which could be useful for mitigating  the  detrimental  effects of magnetism on superconductivity in Majorana devices.


\section{Non-Hermitian Rashba Normal-Superconductor junction}
\label{section4}
\begin{figure}[!t]
	\centering
	\includegraphics[width=0.495\textwidth]{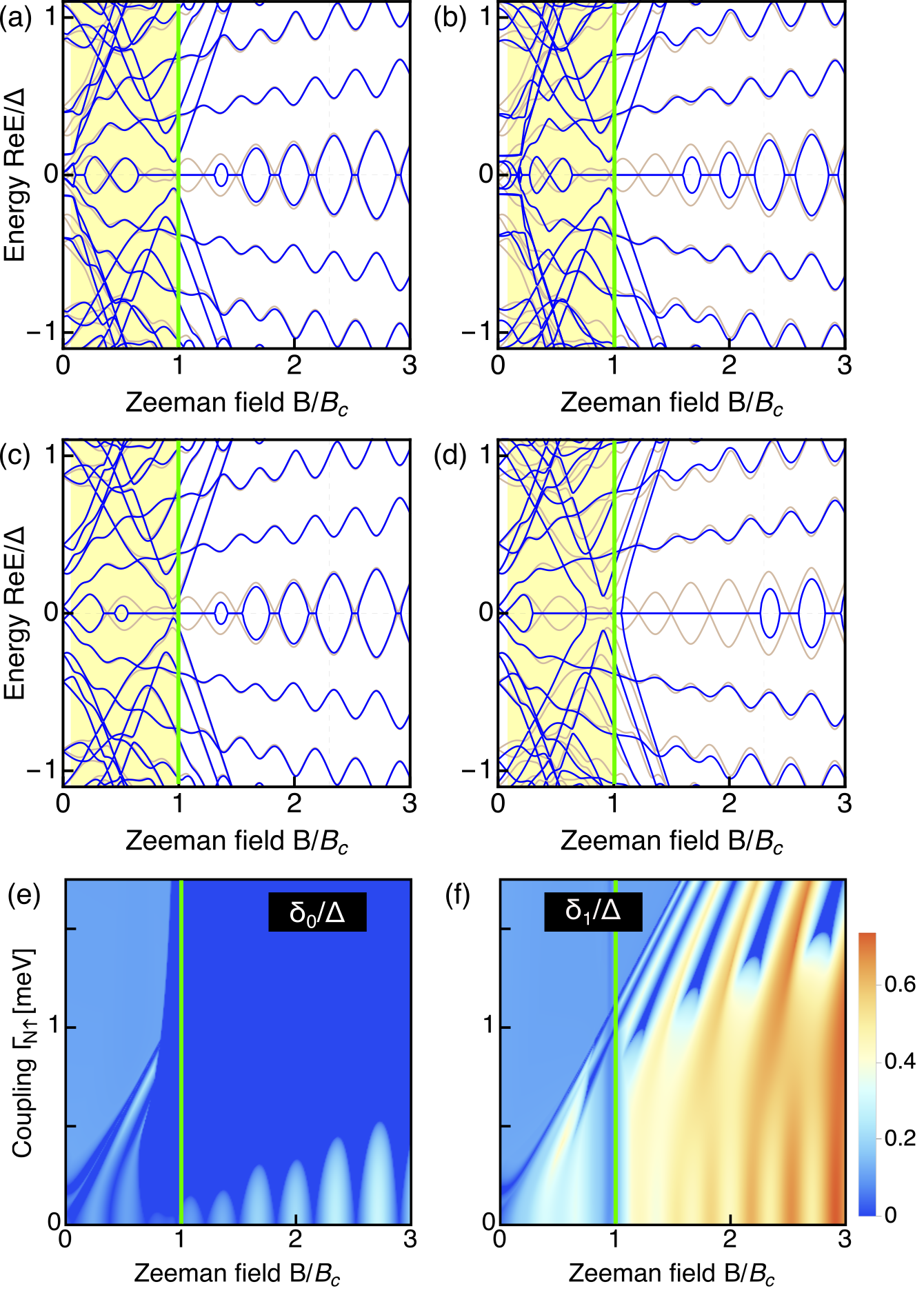} 
	\caption{(a,b) Real low energy spectrum as a function of the Zeeman field for a finite NS Rashba junction with a homogeneous non-Hermiticity in N characterized by $\Gamma_{{\rm N}\uparrow}=0.15$meV (a) and $\Gamma_{{\rm N}\uparrow}=0.25$meV (b). (c,d) The same as in (a,b) but with a homogeneous non-Hermiticity in S characterized by $\Gamma_{{\rm S}\uparrow}=0.1$meV (c) and $\Gamma_{{\rm S}\uparrow}=0.25$meV (d).  The brown curves in (a-d) correspond to the eigenvalues in the Hermitian regime, showing trivial ABSs below the Hermitian topological phase transition at $B=B_{\rm c}$.  The green vertical line in (a-f) marks   $B=B_{\rm c}$, while the shaded yellow regions indicate the Zeeman fields at which trivial ABSs form. (e,f) Lowest and first excited real positive energies $\delta_{0,1}$ as a function of $\Gamma_{{\rm N}\uparrow}$ and $B$.
	  Parameters: $\alpha_{\rm R}=20$meVnm, $\mu_{\rm N}=0.05$meV, $\mu_{\rm S}=0.5$\,meV, $L_{\rm S}=1\mu$m, $L_{\rm N}=1\mu$m, $\Gamma_{{\rm N}\downarrow}=0$, $\Gamma_{{\rm S}\downarrow}=0$.}
	\label{fig4}
\end{figure}

Having discussed the impact of non-Hermiticity on MBSs  emerging in a finite length Rashba superconductor, here we study how non-Hermiticity affects the low-energy spectrum of normal-superconductor (NS) junctions with Rashba SOC. NS junctions in Majorana devices are particularly relevant because they host trivial Andreev bound states (TABSs)    well below the topological phase transition at $B=B_{\rm c}$ by depleting the N region, see Ref.\,\cite{PhysRevB.91.024514,PhysRevB.104.L020501}. The non-Hermitian NS junction is modelled by Eq.\,(\ref{modelTBNW}) but with $\Delta_{n}=0$ in N and $\Delta_{n}=\Delta$ in S, which gives N(S) regions of distinct lengths $L_{\rm N(S)}$.  We consider two independent cases of non-Hermiticity due to homogeneously  coupling the entire N or S region to a ferromagnet lead; the self-energy  due to coupling to the lead is characterized by  $\Gamma_{{\rm N(S)}\sigma}$. The Re low-energy spectrum as a function of the Zeeman field $B$ is presented in Fig.\,\ref{fig4}(a,b) and Fig.\,\ref{fig4}(c,d) for two values of non-Hermiticity in N and S, respectively.  The Im part of the spectrum is not shown because it makes more difficult to analyze the already dense panels.  For completeness we also show the Hermitian low-energy spectrum (brown curves), which exhibits parity crossings with zero-energy loops well below $B_{\rm c}$ (green vertical line) and indicated by the shaded yellow region in Fig.\,\ref{fig4}(a-d).  In Fig.\,\ref{fig4}(e,f) we show the lowest and first excited energy positive Re levels   $\delta_{0(1)}$ as a function of $B$ and $\Gamma_{{\rm N}\uparrow}$, where the blue color indicates $\delta_{0,1}=0$.

The first observation is that in the two situations, with non-Hermiticity in N or S, a zero-energy pinning of both TABSs and MBSs occurs, with similarities but also with some slight differences, see Fig.\,\ref{fig4}(a-d) and also Fig.\,\ref{fig4}(e,f).   Among the similarities is that the zero-energy pinning occurs between  EP transitions: the ends of the zero-energy lines  mark the formation of EPs. We have verified that the Im parts     form loops between  EPs  as those seen in Fig.\,\ref{fig2}(a,b), while they and the associated wavefunctions coalesce at EPs as expected at EPs.  Another similarity is that increasing the strength of non-Hermiticity favors the appearance of longer zero-energy lines between EPs, which corresponds to a zero-energy pinning effect for a larger range of Zeeman fields, as seen by comparing Fig.\,\ref{fig4}(a) and Fig.\,\ref{fig4}(b) or Fig.\,\ref{fig4}(c) and Fig.\,\ref{fig4}(d).  

Among the differences between non-Hermiticity in N and S, we find distinct impact of non-Hermiticity on the lowest energy levels  and also on the  excited energies. For instance the zero-energy crossings in the trivial phase ($B<B_{\rm c}$) and topological phase ($B>B_{\rm c}$) are more susceptible to non-Hermiticity in S than in N. This is seen in Fig.\,\ref{fig4}(a,c) by noting that even a smaller value of non-Hermiticity in S  produces a stronger zero-energy pinning in the trivial phase, see Fig.\,\ref{fig4}(c). The different response  remains even when the strength of non-Hermiticity in N and S are the same, as seen the  larger zero-energy lines in Fig.\,\ref{fig4}(d) as compared to Fig.\,\ref{fig4}(b). Of course that stronger values of non-Hermiticity in N have the potential to produce larger regions with zero-energy pinning but then MBSs acquire zero energy faster than TABSs, see Fig.\,\ref{fig4}(e).  Another difference is that  the excited spectrum remains largely unaffected for reasonable  values of non-Hermiticity in N while the same is not true when non-Hermiticity is in S, see e.g., Fig.\,\ref{fig4}(a,c).   As a result, having non-Hermiticity   in N leads to a gap closing feature occurring at a single point at $B=B_{\rm c}$ and not accompanied by additional states [Fig.\,\ref{fig4}(a,b)]. In contrast, for non-Hermiticity in S, the gap closing can occur at a continuous set of points as a flattened zero-energy line whose ends mark the formation of  EPs [Fig.\,\ref{fig4}(d)]; this is similar to what we saw in Fig.\,\ref{fig2}(a,b) for the gap closing. It is also important to say that stronger non-Hermiticity in N can also affect $\delta_{1}$, inducing it to even vanish either  for   $B<B_{\rm c}$ or $B>B_{\rm c}$, as seen in Fig.\,\ref{fig4}(f). In this case, however,  in the topological phase $\delta_{1}$  vanishes but at stronger non-Hermiticity values than in the trivial phase. Despite the differences, it is clear that non-Hermiticity induces a zero-energy pinning in both MBSs and TABSs.

 \begin{figure}[!t]
	\centering
	\includegraphics[width=0.495\textwidth]{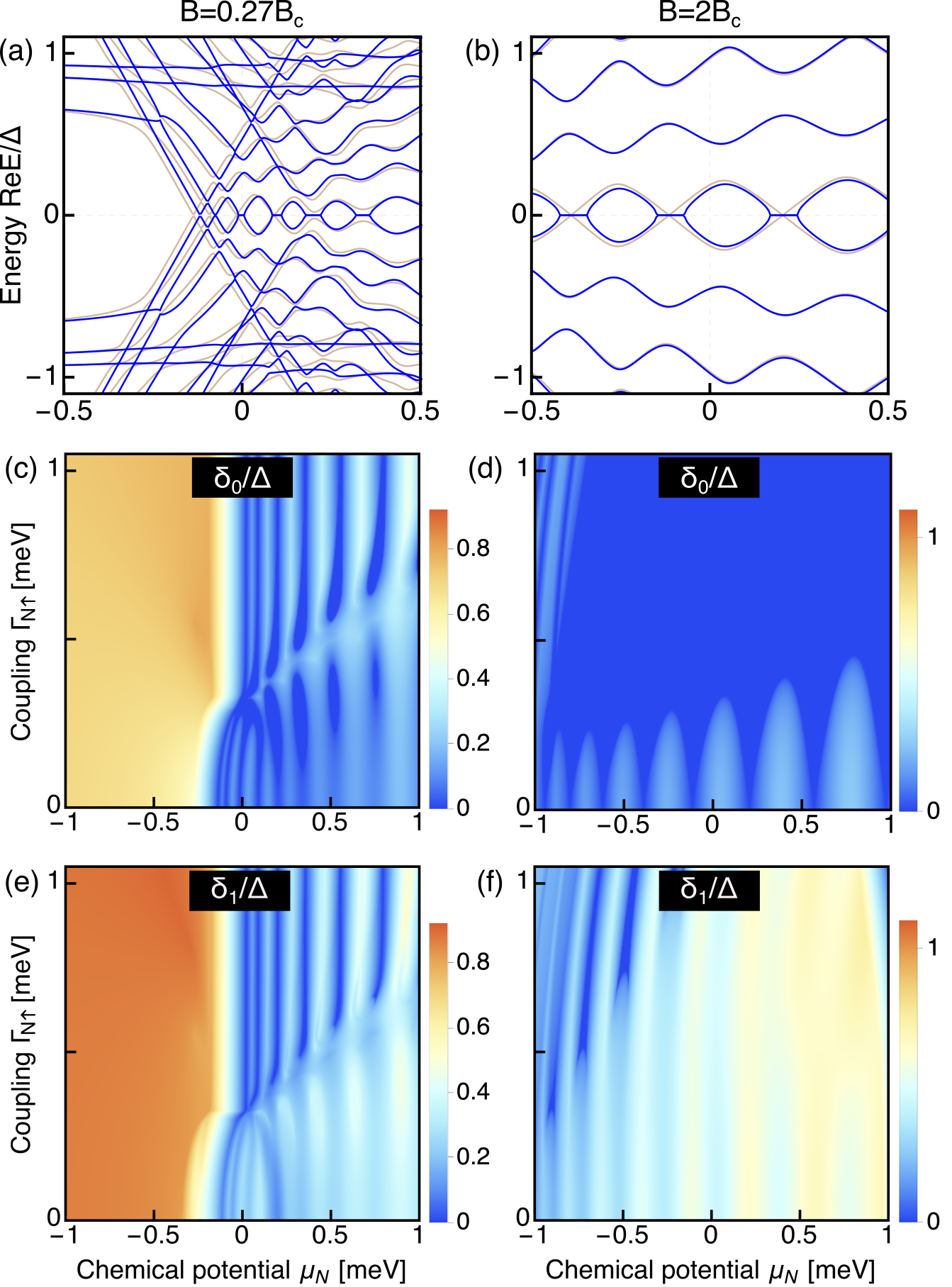} 
	\caption{(a,b) Real low-energy spectrum for a finite NS Rashba junction as a function of the chemical potential in N $\mu_{\rm N}$ at $B$ in the trivial  (a) and topological (b) regimes. Here,   non-Hermiticity is finite and   homogeneous in N and characterized by $\Gamma_{{\rm N}\uparrow}=0.15$meV.  (c,d) Lowest positive real energy $\delta_{0}$ as a function of $\mu_{\rm N}$ and $\Gamma_{{\rm N}\uparrow}$. (e,f) The same as in (c,d) but for the first excited positive real energy $\delta_{1}$.  Parameters: $\alpha_{\rm R}=20$meVnm, $\mu_{\rm S}=0.5$\,meV, $L_{\rm S}=1\mu$m, $L_{\rm N}=1\mu$m, $\Gamma_{{\rm N}\downarrow}=0$.}
	\label{fig5}
\end{figure}

To gain further understanding on the impact of non-Hermiticity on the TABSs and MBSs, in Fig.\,\ref{fig5}(a,b) we plot the Re low-energy spectrum as a function of the chemical potential in N $\mu_{\rm N}$ at $B<B_{\rm c}$ and $B>B_{\rm c}$. In Fig.\,\ref{fig5}(c,d) we plot the lowest Re positive energy $\delta_{0}$ as a function of  the chemical potential $\mu_{\rm N}$ and the coupling $\Gamma_{{\rm N}\uparrow}$, while in Fig.\,\ref{fig5}(e,f) we do the same for the first excited positive Re energy level $\delta_{1}$. As we have done before, in brown color we also show the Hermitian energy levels, which form loops around zero-energy with parity crossings. These oscillatory zero-energy loops reflect the formation of MBSs and TABSs in the topological and trivial phases, respectively, see Ref.\,\cite{PhysRevB.91.024514}.  The first feature we observe is that the parity crossings transform into zero-energy lines with their ends marking EPs, reflecting that the zero-energy pinning   of MBSs and TABSs can be controlled by $\mu_{\rm N}$, see Fig.\,\ref{fig5}(a,b). Although the zero-energy pinning effect is similar in the trivial and topological phases, the impact of non-Hermiticity on the  parity crossings in the topological phase are more likely to give rise to larger zero-energy lines [Fig.\,\ref{fig5}(a,b)]. In fact, by increasing non-Hermiticity via $\Gamma_{{\rm N}\uparrow}$, the trivial parity crossings exhibit a zero-energy pinning but not all of them form zero-energy lines at the same value of non-Hermiticity [Fig.\,\ref{fig5}(c)]. This is in contrast to what occurs for the parity crossings in the topological phase, where all of them simultaneously feel the impact of non-Hermiticity and exhibit a zero-energy pinning [Fig.\,\ref{fig5}(d)]. Furthermore, the energy gap separating the TABSs and MBSs from the quasicontinuum ($\delta_{1}$) is approximately  robust for small values of $\Gamma_{{\rm N}\uparrow}$, but can undergo EP transitions at zero energy when such $\Gamma_{{\rm N}\uparrow}$ is rather strong, as seen in   Fig.\,\ref{fig5}(e,f).

Non-Hermiticity is, therefore, able to  stabilize  both MBSs and TABSs at zero-energy, a zero-energy pinning effect that can be controlled by the Zeeman field or chemical potential of the normal region \footnote{Non-Hermiticity, however, cannot help distinguishing between zero-energy TABSs and zero-energy MBSs. Nevertheless, the distinct features of the spectrum in the trivial (with TABSs) and topological  (with MBSs) phases presented in Figs.\,\ref{fig4} and \ref{fig5} shows that being able to control the amount of non-Hermiticity can be useful to distinguish between TABSs and MBSs. However, this task is challenging because it involves to have good control over the coupling to reservoirs.}. While zero-energy pinning can be beneficial,  it could also bring difficulties  because it will be more challenging to identify the origin of such stable zero-energy states. Moreover, even though reasonable values of non-Hermiticity do not considerably affect the energy gap separating MBSs or TABSs from the quasicontinuum, strong non-Hermiticity can be detrimental.

\section{Conclusions}
\label{section5}
In conclusion, we have investigated the impact of non-Hermiticity on the low-energy spectrum of finite superconducting systems with Rashba spin-orbit coupling, where non-Hermitian effects arise due to coupling to normal or ferromagnet leads.  We have  demonstrated that  non-Hermiticity transforms the Hermitian parity crossings of the oscillatory Majorana energies into  lines of zero real energy whose ends mark the formation of exceptional points. We have  also  found that non-Hermiticity   induces a similar zero-energy pinning effect of trivial Andreev bound states, which  appear well below the topological phase transition in the Hermitian regime. However,  we obtained that Majorana bound states can be more susceptible to non-Hermiticity than trivial Andreev bound states, specially when  non-Hermiticity is present all over the superconductor of  normal-superconductor junctions.  Moreover, we have shown that the values of non-Hermiticity inducing the zero-energy pinning effect do not damage the energy gap that separates the Majorana or Andreev bound states from the quasicontinuum, thus highlighting the beneficial effect of non-Hermiticity. We found that the zero-energy pinning effect can be highly controllable by the interplay of non-Hermiticity and the system parameters, such as Zeeman field and chemical potentials.

We have also revealed that  non-Hermiticity  has an important effect on the Hermitian topological phase transition when its profile is homogeneous all over  the superconductor. In this case we have discovered that the energy gap  undergoes a zero-energy pinning effect due to exceptional points, leading to a flattened gap closing feature unlike the single point Hermitian topological phase transition. This effect suggests that it is possible to achieve a topological phase transition at  Zeeman fields much lower than in the Hermitian regime, which is important because strong Zeeman fields are often seen as detrimental for superconductivity. However, good control over the non-Hermitian mechanism is needed because very strong non-Hermiticity not only has the potential to induce  large zero-energy pinning ranges but it can also destroy the energy gap. Given that   Majorana devices are often coupled to normal leads, where  non-Hermitian effects are intrinsic, our results can be helpful for understanding the possible mechanisms giving rise to zero-energy states.


\section{Acknowledgements}
We thank R. Aguado, M. Sato and Y. Tanaka for insightful discussions. We acknowledge  financial support from the Swedish Research Council  (Vetenskapsr\aa det Grant No.~2021-04121) and the Carl Trygger’s Foundation (Grant No. 22: 2093). The computations  were  enabled by resources provided by the National Academic Infrastructure for Supercomputing in Sweden (NAISS), partially funded by the Swedish Research Council through grant agreement no. 2022-06725.


\bibliography{biblio}

\end{document}